\documentclass[aps,prl,floatfix,twocolumn,showpacs,preprintnumbers,amsmath,amssymb,superscriptaddress]{revtex4-1}
\usepackage{amsmath}
\usepackage{amssymb}
\usepackage{graphicx}
\usepackage{hyperref} 
\usepackage{bm}
\usepackage{epstopdf}
\usepackage{epsfig}

\begin{document}
\title{Intensity equations for birefringent spin lasers}
\author{Gaofeng Xu}
\affiliation{Department of Physics, University at Buffalo, State University of New York, Buffalo, NY 14260, USA}
\author{David Cao}
\affiliation{Department of Physics, University at Buffalo, State University of New York, Buffalo, NY 14260, USA}

\author{Velimir Labinac}
\affiliation{Department of Physics, University of Rijeka, 51000 Rijeka, Croatia}
\author{Igor \v{Z}uti\'c}
\affiliation{Department of Physics, University at Buffalo, State University of New York, Buffalo, NY 14260, USA}
\affiliation{Department of Physics, University of Rijeka, 51000 Rijeka, Croatia}
\begin{abstract}
Semiconductor spin lasers are distinguished from their conventional counterparts by the presence of spin-polarized carriers. The transfer of angular momentum of the spin-polarized carriers to photons provides important opportunities for the operation of lasers. With the injection of spin-polarized carriers, which lead to the circularly polarized light, the polarization of the emitted light can be changed an order of magnitude faster than its intensity. This ultrafast operation of spin lasers relies on a large birefringence, usually viewed as detrimental in spin and conventional lasers. We introduce a transparent description of spin lasers using intensity equations, which elucidate the influence of birefringence on the intensity and polarization modulation of lasers. 
While intensity modulation is independent of birefringence, for polarization modulation an increase in birefringence directly increases the resonant frequency.
Our results for dynamical operation of lasers provide a guide for their spin-dependent response and spintronic applications beyond magnetoresistance. 
\end{abstract}
\maketitle

\vspace{-.2cm}
\subsection{I. Introduction}
\vspace{-.2cm}

Introducing spin-polarized carriers in semiconductors provides both an opportunity to exceed the performance of
best conventional lasers and realize room-temperature spintronic applications, beyond the usual magnetoresistive effects.
While typical spintronic devices rely on unipolar transport: only one type of carriers (electrons) plays an active role,
laser are bipolar devices, a simultaneous description of electrons and holes is crucial~\cite{Chuang:2009,Coldren:2012,Michalzik:2013}.

Spin lasers~\cite{Hallstein1997:PRB,Ando1998:APL,Rudolph2005:APL,Gerhardt2006:EL,Holub2007:PRL,Holub2007:PRL,Hovel2008:APL,Basu2009:PRL,%
Saha2010:PRB,Iba2011:APL,Frougier2013:APL,Cheng2014:NN} embody common elements for spintronic devices: spin injection, relaxation, transport, and detection~\cite{DasSarma2001:SSC,Maekawa:2002,Zutic2004:RMP,Tsymbal:2019,Hirohata2020:JMMM}. This is depicted in Fig.~1(a) for vertical cavity surface emitting lasers (VCSELs) where spin-polarized carriers are injected from magnetic contacts or, alternatively, by using circularly polarized light~\cite{Zutic2020:SSC}. The spin transport is dominated by electrons (bright colors) since the spin imbalance of holes (pale colors) is quickly lost, as they experience stronger spin-orbit coupling and have a much shorter spin relaxation 
time, $\tau_{sp} \ll \tau_{sn} \equiv \tau_s$~\cite{Zutic2004:RMP,Fabian2007:APS,Hilton2002:PRL}. Through the transfer of angular momentum, the spin injection is detected as a circularly polarized light, the photon densities  of positive and negative helicity, $S^+$ and $S^-$, are inequivalent. 

Even though the individual elements of spin laser have been extensively studied~\cite{Zutic2020:SSC}, the interplay between different timescales 
for carrier, spin, and photon dynamics, is far from understood.  For example, unlike in common spintronic devices, where to preserve 
spin information a long spin relaxation time of electrons is desirable~\cite{Tsymbal:2019}, for optimal dynamical operation instead a very short electron 
spin relaxation time is sought~\cite{Lindemann2019:N}.

\begin{figure}[ht]
\centering
\includegraphics*[width=8.6cm]{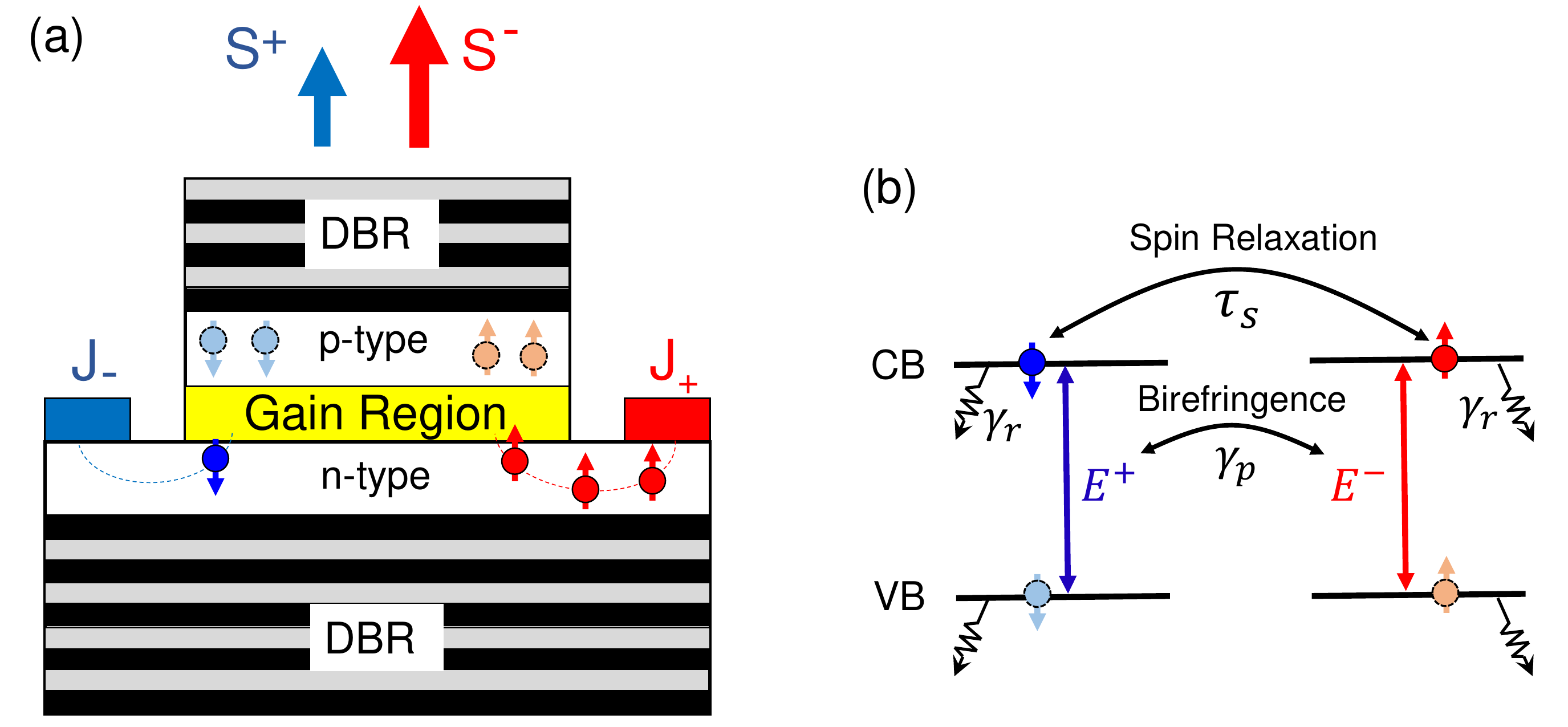}
\vspace{-0.5cm}
\caption{(a) Schematic of a spin laser formed by a gain region, $p$-, $n$-type semiconductor layers, and distributed Bragg reflectors (DBR), with injection of different spins ($J_-<J_+$) and circularly polarized emission  with  photon densities, $S^+<S^-$. (b) Four-level model and carrier-spin-photon dynamics. Carriers in the conduction and valence bands  
(CB, VB, bright and pale colors, respectively) have a recombination rate, $\gamma_r$, spin-relaxation time for electrons is $\tau_s$ and negligible for holes. 
The optical selection rules determine the coupling of spin-down (spin-up) carriers to $E^+$ ($E^-$) light field, while different helicities of light are coupled by 
a linear birefringence, $\gamma_p$.}
\label{fig:Scheme}
\vspace{-0.3cm}
\end{figure}

While many trends in spin lasers can be understood by simply introducing spin-resolved quantities in simple rate equations for 
conventional lasers~\cite{Chuang:2009,Coldren:2012,Michalzik:2013}, this approach leaves large uncertainties for the dynamical 
operation of lasers which can be dominated by optical anisotropies, such as the anisotropy of refractive index--birefringence. To address this situation, 
and motivated by the recent experimental advances showing that a large birefringence with spin injection in III-V quantum well-based lasers supports 
a much faster room-temperature operation than in the best conventional lasers~\cite{Lindemann2019:N}, we introduce here transparent 
intensity equations to elucidate dynamical operation of spin lasers relying on the optical transitions between conduction band (CB) 
and the heavy hole states in the valence band (VB), illustrated in Fig.~1(b). 

An advantage of the intensity equations is their simplicity, instead of the helicity-resolved electric fields with complex amplitudes, $E^\pm$, for the considered 
optical transitions in Fig.~1(b), it is sufficient to use real-valued photon densities, $S^\pm=|E^\pm|^2$. Our approach offers analytical solutions for several situations and  provides a direct link to the extensively studied rate equations for both conventional and spin 
lasers~\cite{Chuang:2009,Coldren:2012,Michalzik:2013,Rudolph2005:APL,Holub2007:PRL,Gothgen2008:APL,Saha2010:PRB,Lee2010:APL,Lee2012:PRB,Lee2014:APL}

These intensity equations are closely related to the spin-flip model~\cite{SanMiguel1995:PRA}, introduced to explain the polarization dynamics in conventional VCSELs and later used 
for describing spin lasers~\cite{SanMiguel1997:IEEE,vanExter1998:PRA,Mulet2002:IEEEJQE,Sande2003:PRA,
Li2010:APL,Gerhardt2011:APL,Al-Seyab2011:IEEEPJ,Gerhardt2012:AOT,Alharthi2015:APL,Lindemann2016:APL,Yokota2018:APL,Adams2018:SST}. 
We show how to correct some of the assumptions in that model, which are particularly important for spin lasers
and their potential to be used for ultrafast operation as a building block of high-performance optical interconnects~\cite{FariaJunior2015:PRB,Lindemann2019:N,Yokota2018:APL},
importing to for a growing need of transferring information~\cite{Hecht2016:N,Miller2017:JLT,Jones2018:N}.
Following this introduction, in Sec.~II we describe our intensity equations. In Sec.~III we introduce dynamic operation of lasers and how it 
is experimentally realized in highly-birefringent spin lasers. In Sec.~IV our results for intensity and polarization modulation are given, and in Sec.~V 
we  provide conclusions  and note some open questions for future work.

\section{II. Intensity Equations}

The polarization dynamics of VCSELs has been successfully described by the influential spin-flip model (SFM)~\cite{SanMiguel1995:PRA}
and widely applied to conventional lasers, having no external source of spin-polarized 
carriers~\cite{SanMiguel1997:IEEE,vanExter1998:PRA,Mulet2002:IEEEJQE,Sande2003:PRA}.
For a spin laser the corresponding equations can be generalized by including injection of spin-polarized carriers
as shown in Fig.~\ref{fig:Scheme}(a). Since the hole spin relaxation is typically much faster than for electrons, 
there is no depicted spin imbalance in the $p$-region~\cite{Zutic2004:RMP}. 

Following the conservation of angular momentum and the optical selection rules~\cite{Zutic2004:RMP}, Fig.~\ref{fig:Scheme}(b) 
illustrates the SFM which focuses on the gain region based on a quantum well (QW) where its confinement splits the heavy and
light hole degeneracy.  In the resulting equation it is then sufficient to consider optical transition between the 
CB,  with the projection of the
total angular momentum ${\bf J}_z =\pm 1/2$ and the 
VB with ${\bf J}_z= \pm 3/2$ for heavy holes,
\begin{eqnarray}
\dot E^{\pm}&=& \frac{1+i\alpha}{2 \tau_{ph}}(N\pm n-1) E^{\pm} -(\gamma_a+i \gamma_p ) E^{\mp},
\label{eq:SFME}
\\
\dot N&=&\gamma_r \left[J_+ (t) +J_- (t)\right]-\gamma_r N -\gamma_r (N+n)\vert E^{+}\vert^2 \nonumber \\
&&-\gamma_r (N-n)\vert E^{-}\vert^2, \\
\label{eq:SFMN}
\dot n&=&\gamma_r\left[J_- (t) -J_+ (t)\right]-n/\tau_s-\gamma_r (N+n)\vert E^{+}\vert^2 \nonumber \\
&&+\gamma_r(N-n)\vert E^{-}\vert^2,
\label{eq:SFMn}
\end{eqnarray}
where the normalized (see Appendix) circularly polarized components of slowly varying amplitudes of the electric field are 
related to linear modes by $E^{\pm}=(E_x \pm i E_y)/\sqrt{2}$. 
Corresponding photon densities are $S^{\pm}=|E^{\pm}|^2$, with a photon lifetime $\tau_{ph}$.  $N$ is the total number of carriers with a recombination rate $\gamma_r$, $n$ is the population difference between spin-down and spin-up electrons with a spin relaxation lifetime $\tau_s$, 
and $\alpha$ is the linewidth enhancement factor. $\gamma_a$ and $\gamma_p$ are the 
dichrosism and linear birefringence, the amplitude and phase anisotropies of the cavity. $J_{\pm} (t)$ is the time-dependent injection rate of spin-up ($+$) and spin-down ($-$) carriers. 
 
The SFM equations contain complex amplitudes of the electric field, which can be expressed in terms of real quantities as $E_{x, y}= {\cal E}_{x, y} \exp (i \phi_{x, y})$. 
Therefore, the equations can be rewritten in terms of the dimensionless real quantities, such that all the frequencies are scaled to $\gamma_r$ and differentiation 
expressed with respect to dimensionless time, $\tau=\gamma_r t$, as
\begin{flalign}
&\dot {\cal E}_x=\left[\frac{N-1}{2 \tau_{ph}}-\gamma_a \right] {\cal E}_x - \frac{n}{2 \tau_{ph}}(\alpha \cos \phi-\sin\phi) {\cal E}_y,  \label{eq:Rx}    &\\
&\dot {\cal E}_y=\left[\frac{N-1}{2 \tau_{ph}}+\gamma_a \right]{\cal E}_y + \frac{n}{2 \tau_{ph}}(\alpha \cos \phi+\sin\phi) {\cal E}_x, \label{eq:Ry}    &\\
&\resizebox{.88\hsize}{!}{$\dot \phi=-2\gamma_p + \frac{n}{2 \tau_{ph}} \left[ \alpha \sin\phi \frac{{\cal E}^2_y-{\cal E}^2_x}{{\cal E}_x {\cal E}_y} +
\cos\phi \frac{{\cal E}^2_x+{\cal E}^2_y}
{{\cal E}_x {\cal E}_y} \right]$,} \label{eq:phi} & \\
&\dot N=J-N\left(1+{\cal E}_x^2+{\cal E}_y^2\right)-2\sin\phi n{\cal E}_x{\cal E}_y, 
\label{eq:N}          &\\
&\resizebox{.88\hsize}{!}{$\dot n=(J_- - J_+)- \frac{n}{\tau_s} -2 \sin\phi N {\cal E}_x{\cal E}_y-n\left({\cal E}_x^2+{\cal E}_y^2\right)$,} \label{eq:n} &
\end{flalign}
where $\phi= \phi_x - \phi_y$ is the phase difference between the two linear modes and $J=J_+ + J_-$ is the total injection.

\subsection{A. Intensity equations without spin injection}
In the absence of spin injection, $J_+= J_-$, the spin polarization of carriers is very small, i.e., $n$ is negligible. 
Therefore, the time evolution of the phase can be approximated, using dimensionless time $\tau=\gamma_r t$, by 
\begin{equation}
\phi \approx -2 \gamma_p \tau.
\label{eq:phi_approx}
\end{equation}
Considering typically small spin relaxation times in semiconductors used in gain region of a laser~\cite{Lindemann2016:APL,Lindemann2019:N}, 
$1/\tau_s\gg \gamma_r$, we can adiabatically eliminate $n$ ($\dot n \approx 0$) to obtain
\begin{equation}
n \approx -2\tau_s  \sin\phi N {\cal E}_x {\cal E}_y.
\label{eq:n_approx}
\end{equation}
With the approximations in Eqs.~(\ref{eq:phi_approx}) and (\ref{eq:n_approx}), the SFM from Eqs.~(\ref{eq:SFME})-(\ref{eq:SFMn})
is reduced to dynamic equations for the light intensities $S_{x, y} = {\cal E}_{x, y}^2$ and total carrier number $N$
\begin{eqnarray}
\dot S_x &=&S_x\left[  (N-1)/\tau_{ph}- 2\gamma_a  -\epsilon_{xy}S_y \right], 
\label{eq:Sx_3} \\
\dot S_y &=&S_y\left[ (N-1) /\tau_{ph}+2\gamma_a-\epsilon_{yx}S_x  \right],  
\label{eq:Sy_3} \\
\dot N &=&J-N -N(S_x+S_y)+2 \tau_s N S_x S_y, 
\label{eq:N_3}
\end{eqnarray}
where the cross-saturation coefficients are $\epsilon_{xy}=\epsilon_{yx} =\tau_s/\tau_{ph}$ which suppress the intensity of the emitted light
as the carrier injection is increased. However, the above equations arising from the SFM, lack the well-known self-saturation effects
in conventional lasers known to be crucial in limiting the intensity of the emitted light at large injection levels~\cite{Chuang:2009,Coldren:2012,Lee1993:OQE} 
and also studied in the rate-equation description of spin lasers~\cite{Holub2007:PRL,Gothgen2008:APL}. 

For a more complete description of the gain saturation (also referred to as the gain compression), we phenomenologically introduce self-saturation 
terms with coefficients $\epsilon_{xx}$ and $\epsilon_{yy}$ for the $x$ and $y$ modes 
\begin{eqnarray}
\dot S_x &=&S_x\left[  (N-1) /\tau_{ph}- 2\gamma_a  -\epsilon_{xy}S_y-\epsilon_{xx} S_x \right],  
\label{eq:Sx_3SAT}\\
\dot S_y &=&S_y\left[ (N-1)/\tau_{ph}+2\gamma_a-\epsilon_{yx}S_x-\epsilon_{yy} S_y \right],  
\label{eq:Sy_3SAT} \\
\dot N &=&J-N -N(S_x+S_y)+2 \tau_s N S_x S_y,
\label{eq:N_3SAT}
\end{eqnarray}
where we note that in describing conventional lasers the gain saturation coefficients are often simply 
given by $\epsilon_{xx}=\epsilon_{yy}=\epsilon$ and  $\epsilon_{xy}=\epsilon_{xy}=0$~\cite{Chuang:2009,Coldren:2012,Lee1993:OQE}.   

\subsection{B. Intensity equations with spin injection}

The immediate effect of a spin injection, $J_+ \neq J_- $, is a significant spin polarization of carriers, such that 
\begin{equation}
n \approx \tau_s( J_- - J_+) -2\tau_s N R_x R_y \sin\phi,
\label{eq:n_spin}
\end{equation}
which in turn leads to additional terms in the equations for intensities and phase
\begin{eqnarray}
\dot S_x &=&S_x\left[ (N-1)/\tau_{ph}- 2\gamma_a  -\epsilon_{xy}S_y-\epsilon_{xx} S_x \right] 
\label{eq:Sx_SAT}
\\
&&-\sqrt{1+\alpha^2} \frac{\tau_s}{\tau_{ph}} (J_- - J_+)\sqrt{S_x S_y} (\alpha \cos \phi -\sin \phi),  \nonumber \\
\dot S_y &=&S_y\left[(N-1)/\tau_{ph}+2\gamma_a-\epsilon_{yx}S_x-\epsilon_{yy} S_y \right] 
\label{eq:Sy_SAT}\\
&&+\sqrt{1+\alpha^2}  \frac{\tau_s}{\tau_{ph}} (J_- - J_+) \sqrt{S_x S_y} (\alpha \cos \phi +\sin \phi), \nonumber \\ 
\dot N &=&-N+(J_+ + J_-)-N(S_x+S_y)+2 \tau_s N S_x S_y,  
\label{eq:N_SAT}
\end{eqnarray}
\vspace{-0.7cm}
\begin{eqnarray}
\begin{split}
\dot \phi = & -2\gamma_p +n/(2\tau_{ph})  \left[ \alpha \sin \phi \left(\sqrt{S_y/S_x} - \sqrt{S_x/S_y} \right)  \right.    \\
 & + \left. \cos \phi \left(\sqrt{S_y/S_x} + \sqrt{S_x/S_y} \right) \right]. \\
\label{eq:phi_SAT}
\end{split}
\end{eqnarray}

\vspace{-0.5cm}
The above Eqs.~(\ref{eq:Sx_SAT})-(\ref{eq:phi_SAT}), with real-valued quantities, can be used to study the 
dynamic operation of spin lasers and provides a good agreement with the common SFM~\cite{SanMiguel1995:PRA}, 
as shown in Appendix. 
The transparency of this approach allows analytical solutions of intensity modulation 
response by a small-signal analysis and offers opportunities to further explore the dynamics of highly-birefringent lasers
using linear analysis.

\section{III. Dynamic Operation}

\begin{figure}[ht]
\centering
\includegraphics*[width=8.6cm]{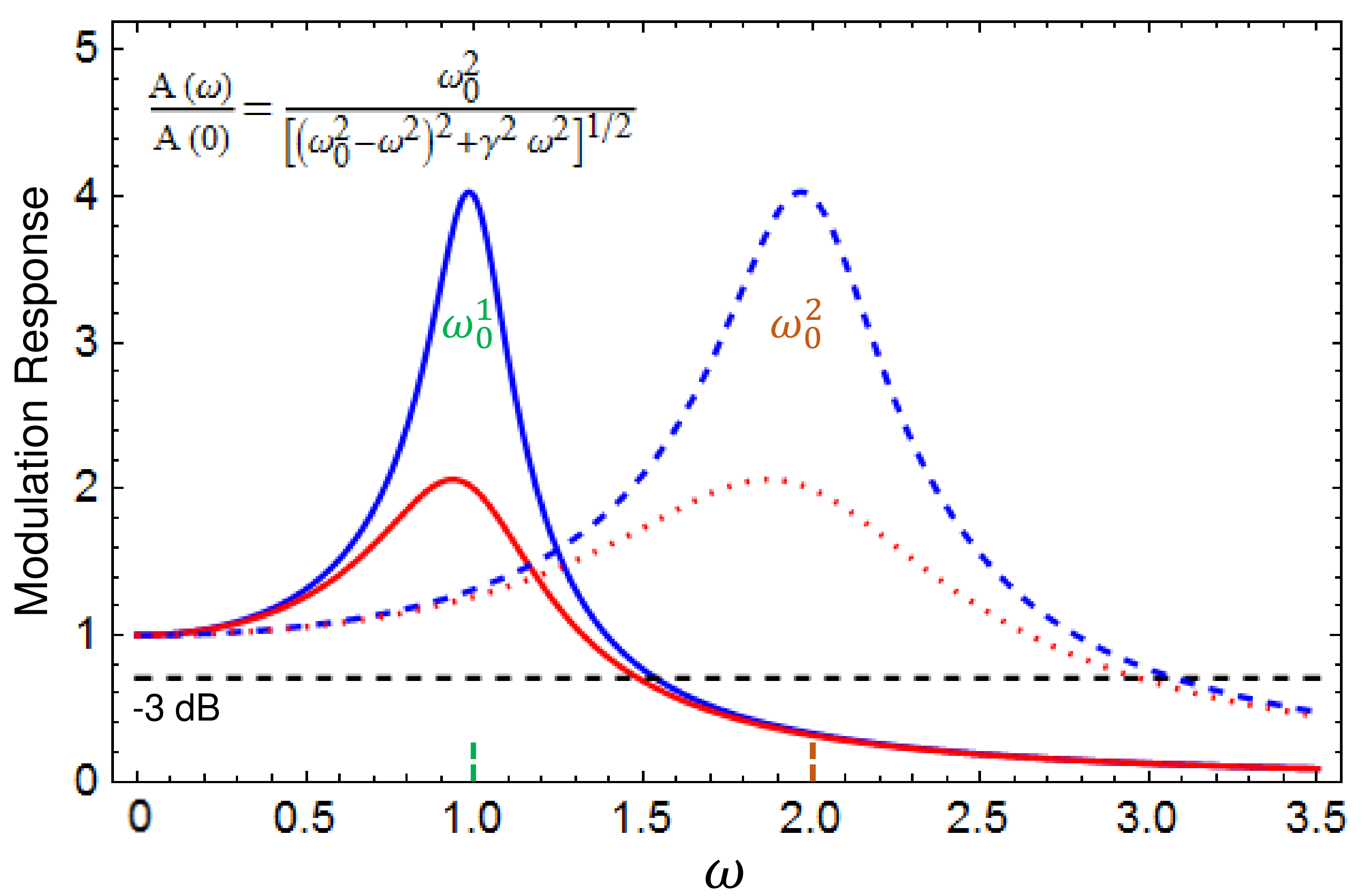}
\vspace{-0.5cm}
\caption{(a) Modulation response, characterized by the normalized amplitude $A(\omega)/A(0)$, of a driven damped harmonic oscillator with natural frequencies $\omega_0^{1,2}$ and damping rates $\gamma=\omega_0^{1,2}/2,  \omega_0^{1,2}/4$. The dashed horizontal line indicates $-3$ dB level as a limit for significant response.
}
\label{fig:HO}
\vspace{-0.3cm}
\end{figure}

The most attractive properties of conventional lasers usually lie in their dynamical performance, suitable for transferring information and 
implementing optical interconnects~\cite{Chuang:2009,Coldren:2012,Michalzik:2013}. A damped driven harmonic oscillator, 
${\ddot x}+\gamma {\dot x}+\omega^2_0x=(F_0/m) \cos\omega t$,
provides a valuable model for the dynamic operation of lasers~\cite{Lee2012:PRB}, 
where $\omega_0$ is the  angular frequency of the simple harmonic 
oscillator, $\gamma$ is the damping constant, $F_0$ is the amplitude 
of the driving force and $m$ is the mass. 

Such a harmonic oscillator shares with lasers its resonant behavior near the angular frequency 
$\omega \approx \omega_0$ and a large reduction of the amplitude, $A(\omega)$, for $\omega \gg \omega_0$, as depicted for two resonant
frequencies in Fig.~\ref{fig:HO},
\begin{equation}
A(\omega)/A(0)=\omega_0^2/\left[(\omega_0^2-\omega^2)^2+\gamma^2\omega^2\right]^{1/2}.
\label{eq:HO}
\end{equation}
The reduction of  $A(\omega)$ by -3 dB, compared to $A(0)$,  gives a useful frequency range over which
substantial signals can still be transferred, corresponding to the modulation bandwidth of a laser~\cite{Chuang:2009,Lee2012:PRB}. 

A challenge for spin lasers is to seek improving dynamic operation over their best conventional counterparts. 
Already the first VCSEL with optical spin injection~\cite{Hallstein1997:PRB} has supported a high-frequency operation. 
The transfer of a Larmor precession of the electron spin to the spin of photons was shown polarization oscillation of the emitted light up to 
44 GHz in a magnetic field of 4 T at 15 K~\cite{Hallstein1997:PRB}. While this approach is limited to cryogenic temperatures and does not 
allow an arbitrary modulation of the polarization, needed for high-speed information transfer, nor it is clear if the
resulting modulation bandwidth (recall Fig.~\ref{fig:HO}) could exceed those from conventional semiconductors, 
it has stimulated subsequent studies in spin lasers. 
 
\begin{figure}[ht]
\centering
\includegraphics*[width=8.6cm]{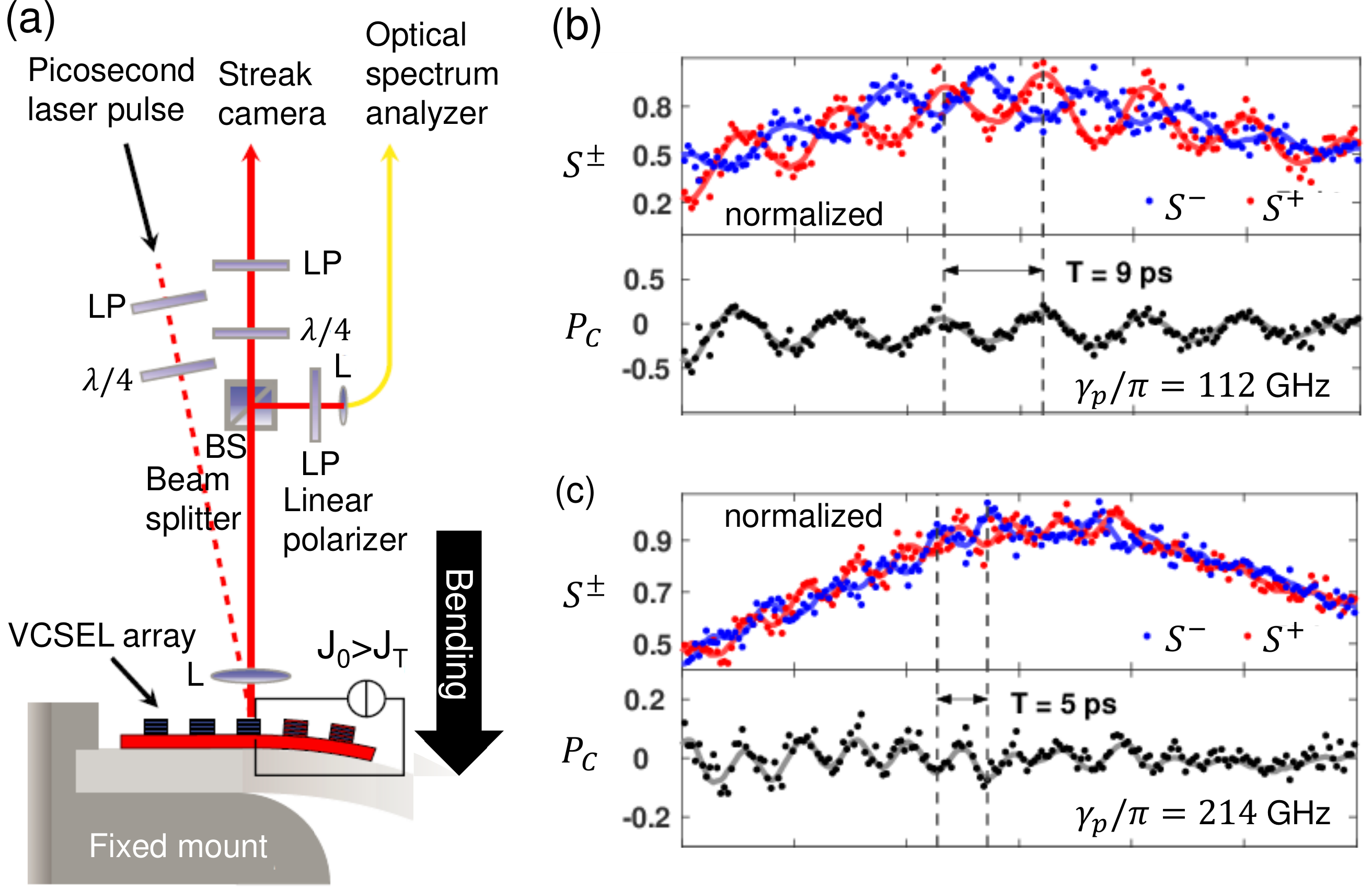}
\vspace{-0.5cm}
\caption{(a) Experimental detection of the polarization dynamics of a spin laser pumped by a constant electrical injection $J_0$ above the threshold $J_T$ and a circularly polarized ps laser pulse as spin injection. Birefringence, $\gamma_p$, is controlled  by a mechanical bending on the VCSEL array, which induces cavity anisotropy. The setup contains linear polarizers (LP), quarter-wave plates ($\lambda/4$), a beam splitter (BS), and lenses (L). The laser output is detected by a streak camera and an optical spectrum analyzer. (b), (c) Polarization dynamics of the laser after a pulsed spin injection for $\gamma_p/\pi$ $112$ and $214$ GHz. $S^{\pm}$ are the helicity-resolved light intensities, $P_C$ 
is the circular polarization degree, Eq.~(\ref{eq:PC}), and $T$ denotes the period of the polarization oscillation. 
From Ref.~\cite{Lindemann2019:N}. 
}
\label{fig:Experiment}
\vspace{-0.3cm}
\end{figure}

One such realization of spin lasers supporting room-temperature ultrafast operation was demonstrated in highly-birefringent VCSELs~\cite{Lindemann2019:N}, as shown in Fig.~\ref{fig:Experiment}. The role of birefringence can be understood from Fig.~\ref{fig:Scheme}(b) and SFM or intensity equations from Sec.~II. Since the birefringence is responsible for the beating between the emitted light of different helicities,  the changes in the polarization of the emitted light, 
\begin{equation}
P_\textrm{C}=(S^+-S^-)/(S^++S^-),
\label{eq:PC}
\end{equation}
can be faster than the changes in the light intensity. While initially these polarization changes were limited to $\sim$ 10 GHz for commercial III-V VCSELs
to which spin-polarized carriers were optically injected~\cite{Gerhardt2011:APL,Li2010:APL}, subsequent theoretical predictions of much 
higher strain-enhanced birefringence values~\cite{FariaJunior2015:PRB} and their experimental realization~\cite{Lindemann2016:APL} have paved 
the way for spin lasers that could operate faster than the best conventional counterparts. Specifically, the realization of higher birefringence values, 
using elasto-optic effect up to $\sim 80$ GHz~\cite{Panajotov2000:APL},  asymmetric heating up to $\sim 60$ GHz~\cite{Pusch2017:APL},
integrated surface gratings up to 98 GHz~\cite{Pusch2019:EL}, and mechanical bending reaching 259 GHz~\cite{Pusch2015:EL}, by itself only
supports static implications of birefringence due to mode splitting in VCSEL. However, Fig.~\ref{fig:Experiment} also reveals that high birefringence
is also compatible with ultrafast oscillations in $P_C$.

To study the dynamic operation of spin lasers,  with spin polarization of injected carriers
\begin{equation}
P_J = (J_+-J_-)/ (J_++J_-),
\label{eq:PJ}
\end{equation}
and conventional lasers as their special limiting case, where $P_J\equiv 0$, 
it is convenient that each of  the key quantities,  $X$ (such as, $J$, $S$, $N$ and $P_J$), is decomposed into a steady-state $X_0$ and a modulated part
$\delta X(t)$~\cite{Lee2012:PRB}, 
\begin{equation}
X=X_0+\delta X(t), 
\label{eq:X}
\end{equation}
where we can assume harmonic modulation $\delta X(t)=\operatorname{Re}[\delta X(\omega) e^{-i \omega t}]$.

\begin{figure}[ht]
\centering
\includegraphics*[width=8.6cm]{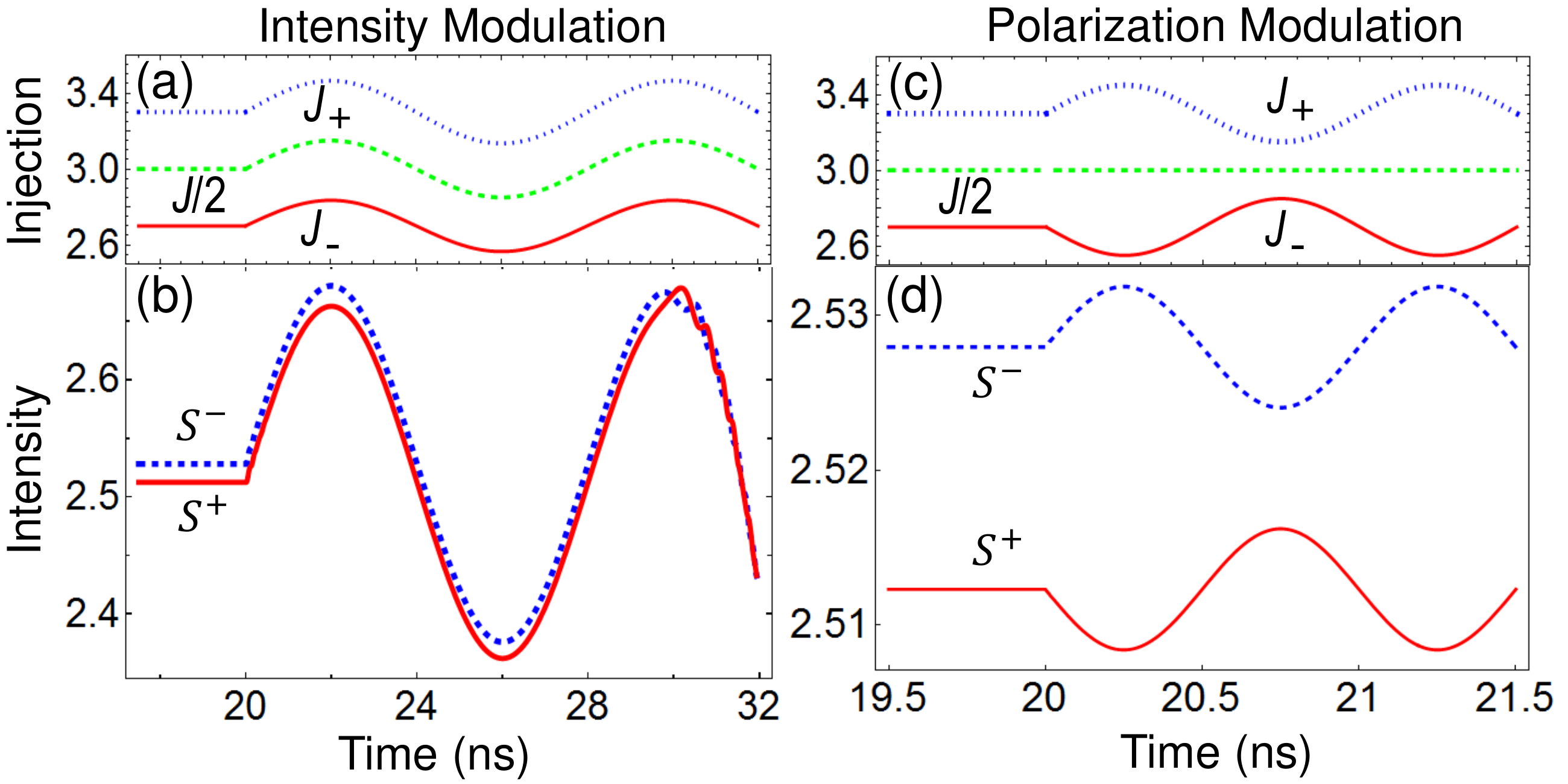}
\vspace{-0.5cm}
\caption{Time-dependence of the spin injection $J_{\pm}$ and helicity-resolved light intensities $S^{\pm}$ for intensity  [(a),  (b)] and polarization modulation [(c), (d)] 
in a spin laser. Before the modulation is turned on at $t=20$ ns,  the total injection $J=J_+ + J_-$ is constant with a spin polarization $P_{J0}=0.1$. 
}
\label{fig:IPM}
\vspace{-0.3cm}
\end{figure}

We focus on the intensity and polarization modulation ({\em IM}, {\em PM}), illustrated in Fig.~\ref{fig:IPM}. {\em IM} for a steady-state polarization implies $J_+ \neq J_-$ (unless $P_J=0$),
\begin{equation}
IM: \:
J=J_0+\delta J \cos(\omega t), \quad  P_J = P_{J0},
\label{eq:AM}
\end{equation}
where $\omega$ is the angular modulation frequency.
Such a modulation can be contrasted with {\em PM}
which also has $J_+ \neq J_-$, but
$J$ remains constant~\cite{const},
\begin{equation}
PM: \:
J=J_0, \quad P_J=P_{J0}+\delta P_J \cos(\omega t).
\label{eq:PM}
\end{equation}

In spin lasers it is also possible to consider other modulation schemes with $P_J\neq0$. For example, a complex modulation~\cite{Boeris2012:APL}
can suppress an undesired frequency modulation, or chirp, a direct consequence of {\em IM} and the carrier dependence of the refractive index in the gain region. In addition to faster operation, by modulating the polarization of the emitted light rather than its intensity~\cite{Lindemann2019:N,Hopfner2014:APL}, 
spin lasers offer a reduced noise and an improved signal transfer~\cite{Wasner2015:APL}.

\section{IV. Intensity and polarization modulation response}

The modulation response in conventional lasers, typically realized using  {\em IM}, can be simply summarized
by relating their resonant (relaxation-oscillation) frequency  $f_R=\omega_{R}^{IM}/2\pi$ and the resulting usable frequency range 
given by the modulation bandwidth~\cite{Chuang:2009,Coldren:2012} (see Fig.~2),
\begin{equation}
f_\mathrm{3dB} \approx \sqrt{1+\sqrt{2}} f_R.
\label{eq:3dB}
\end{equation}
The modulation bandwidth can be estimated by the resonant frequency, $f_R=(1/2\pi)\sqrt{g_0 S_0/[\tau_{ph}(1+\epsilon S_0)]}$,
where $g_0$ is the gain constant, $S_0$ is the steady-state photon density, $\tau_{ph}$ the photon lifetime, used also in the SFM,
and $\epsilon$ is the simplified parameterization of the gain saturation, noted in Sec.~IIA. To enhance the bandwidth one can seek
to enhance $f_R$ by materials design to enlarge $g_0$, decrease $\tau_{ph}$ by reducing the reflectivity of mirrors forming the 
resonant cavity (recall Fig.~1), or by increasing $J$ to attain a larger $S$.  While the last approach is the most common, 
we can see that it comes not only at the cost of the higher-power consumption, but that a finite $\epsilon$ is responsible for the 
saturation of $S$ as $J$ is increased. 

However, this common analysis using Eq.~(\ref{eq:3dB}) excludes the influence of birefringence, which experimentally can exceed $250$ GHz~\cite{Pusch2015:EL}, 
and, even for conventional lasers with $P_J=0$, it is unclear what would be its influence on $f_R$ and the corresponding modulation bandwidth. 
In spin lasers the situation is further complicated as the birefringence can be viewed as undesirable and there efforts in designing lasers to minimize 
it~\cite{Hovel2008:APL,Frougier2015:OE,Yokota2017:IEEEPTL,Fordos2017:PRA}.

To elucidate the role of birefringence of the modulation response we analyze the dynamic operation of the laser using a perturbative approach to the steady-state response, using a decomposition as in Eq.~(\ref{eq:X}), known also as the small signal analysis (SSA)~\cite{Chuang:2009,Coldren:2012}, limited to a small modulation.  This approach is  readily generalized for spin lasers~\cite{Lee2010:APL},  with 
$|\delta J/J_0| \ll 1$ for {\em IM} and $|\delta P_J| \ll 1,  |P_{J0}\pm \delta P_J| \leq1$ for {\em PM}.

From the intensity equations we can obtain $\delta S^{\pm}(\omega)$ and the 
(modulation) frequency response functions 
\begin{equation}
R_\pm(\omega)=|\delta S^\mp(\omega)/\delta J_\pm(\omega)|.
\label{eq:response}
\end{equation}
For $P_J=0$ they reduce to, 
$R(\omega)=|\delta S(\omega)/\delta J(\omega)|$, usually normalized to its $\omega=0$ value, just as in Eq.~(\ref{eq:HO}),
\begin{equation}
\left|  R(\omega)/R(0) \right| = \omega_R^2/\left[(\omega_R^2-\omega^2)^2+\gamma^2\omega^2\right]^{1/2},
\label{eq:band}
\end{equation}
where, 
$\omega_R$ and damping rate $\gamma$ can be analytically extracted from Eq.~(\ref{eq:Sx_3SAT})-(\ref{eq:N_3SAT}).
For example, assuming $S_y=0$, we can obtain the steady-state value, $S_{x0}=J_0/N_0-1$ and $N_0=1-2 \tau_{ph}\gamma_a + \tau_{ph} \epsilon_{xx}S_{x0}$, and conclude that the normalized threshold values are
\begin{equation}
J_T=N_T=1-2\tau_{ph} \gamma_a.
\label{eq:JNT}
\end{equation}
We can then express
\begin{eqnarray}
\omega_R^2&=&(J_0/N_0-1)(N_0/\tau_{ph}+\epsilon_{xx} J_0/N_0), \label{eq:omegaR} \\
\gamma&=&(J_0/N_0)(1+\epsilon_{xx})-\epsilon_{xx},
\label{eq:gamma}
\end{eqnarray}
while assuming instead $S_x=0$, $\omega_R$, and $\gamma$ would retain the same form, but with $\epsilon_{xx}\rightarrow \epsilon_{yy}$.

\begin{figure}[ht]
\centering
\includegraphics*[width=8.6cm]{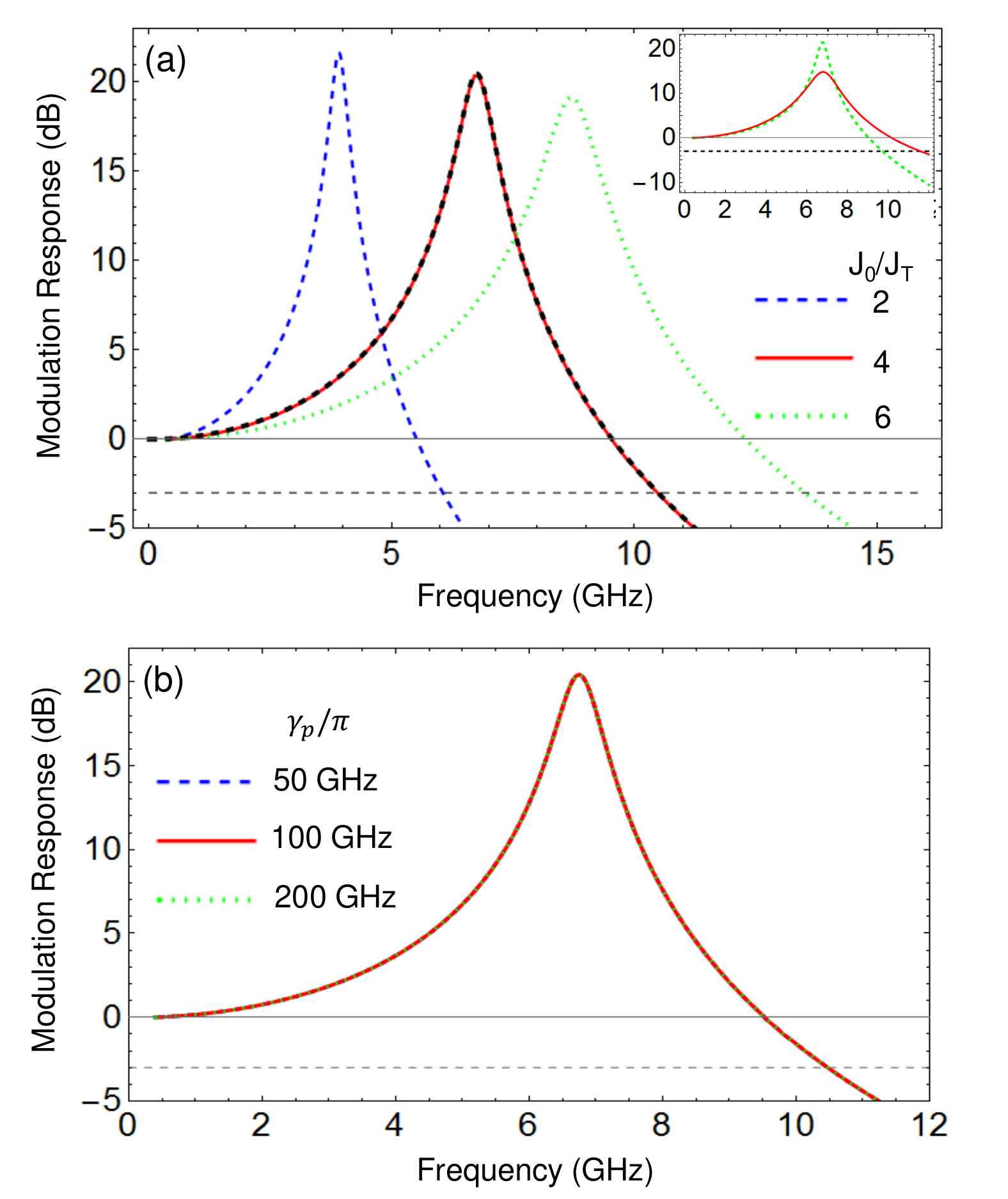}
\vspace{-0.5cm}
\caption{Effects of injection (a) and birefringence (b) on the intensity modulation response. (a) The intensity modulation bandwidth  is enhanced by larger injection, for $J_0/J_T=2$, 4 and 6. 
For $J_0/J_T=4$, the analytical small signal analysis solution of the modulation response is given by the black
curve, showing a good agreement with the numerical result with $\gamma_p/\pi = 100$. 
The inset shows  the bandwidth enhancement by spin polarization of injections from $P_{J0}=0.3$ (green) to $0.9$ (red). 
(b) The intensity modulation response coincides for different birefringence $\gamma_p/\pi = 50$, 100, 200 GHz. 
As in Fig. ~2, the dashed horizontal line indicates $-3$ dB level as a limit for significant response.
}
\label{fig:5}
\vspace{-0.3cm}
\end{figure}

To illustrate the effects of injection and birefringence on {\em IM} 
 explicitly, we calculate the modulation response for a series of injection and birefringence. As shown in Fig.~\ref{fig:5}, the resonant frequency $\omega_R$ as well as bandwidth increase with larger injection, which is also implied by Eqs.~(\ref{eq:3dB}) and (\ref{eq:omegaR}). Note that there is a good agreement between the numerical calculation and analytical expressions in Eqs.~(\ref{eq:band}), (\ref{eq:omegaR}),  (\ref{eq:gamma}). 
Additionally, the {\em IM} bandwidth can be enhanced by increasing the polarization of injection $P_{J0}$, without visibly 
altering the resonant frequency, as shown in the inset of Fig.~\ref{fig:5}(a)
Using the rate equations (in the absence of birefringence) such as increase in $P_{J0}$ has enhanced both the bandwidth and the resonant frequency~\cite{Lee2010:APL,Banerjee2011:JAP}.
In contrast, from Fig.~\ref{fig:5}(b), {\em IM} response is unaffected by birefringence. This can be understood from the intensity equations [Eqs.~(\ref{eq:Sx_SAT})-(\ref{eq:phi_SAT})], in which birefringence only changes the phase difference $\phi$ between $x$ and $y$ modes, rather than the intensities.

\begin{figure}[ht]
\centering
\includegraphics*[width=8.6cm]{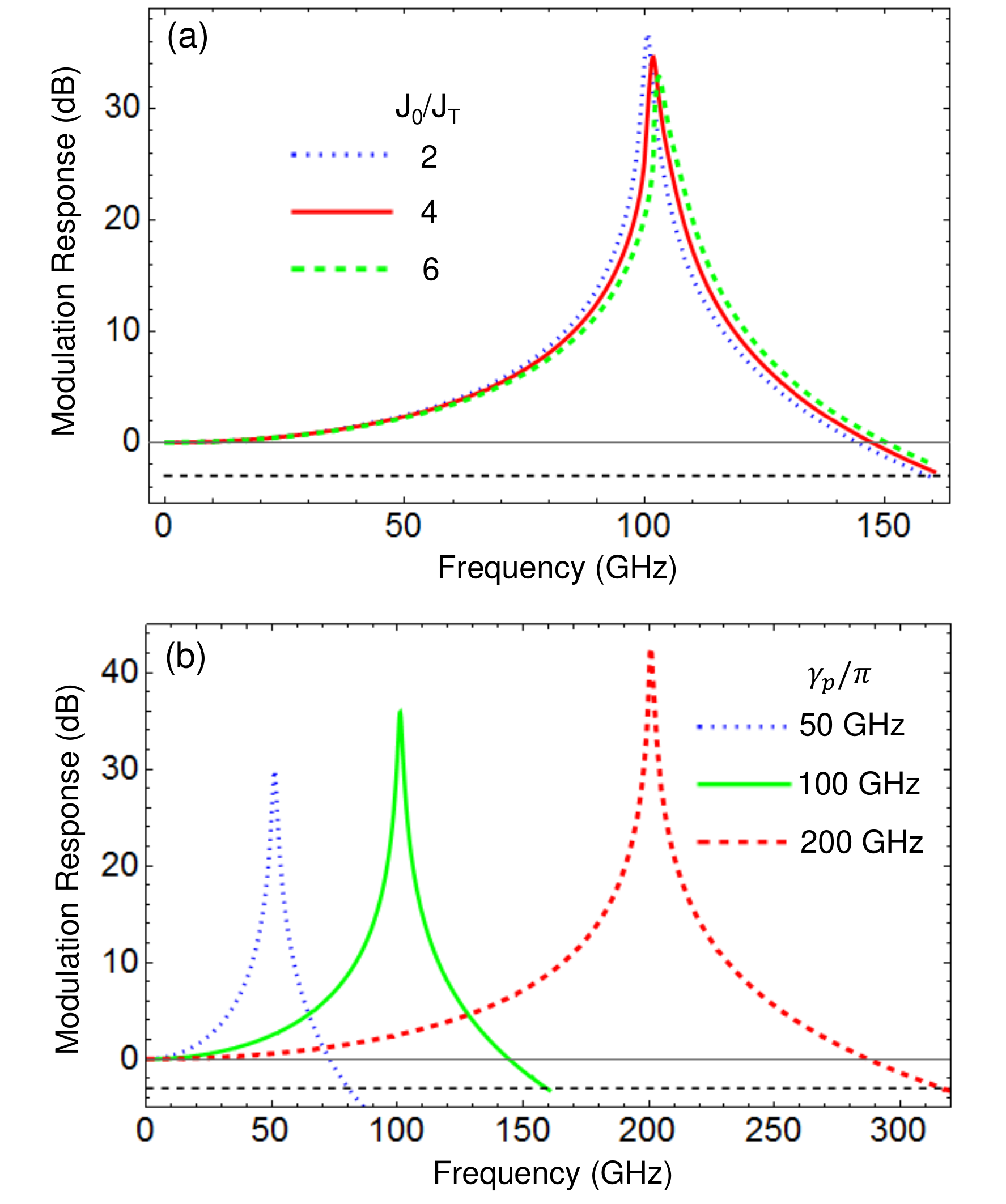}
\vspace{-0.5cm}
\caption{Effects of injection (a) and birefringence (b) on the polarization modulation response. (a) A minor increase in the resonant frequency and bandwidth of polarization modulation for injection $J_0/J_T$  from $2$, $4$ to $6$. Here $\gamma_p/\pi = 100$ GHz. (b) A significant enhancement of resonant frequency and bandwidth with birefringence $\gamma_p/\pi = 50$ GHz, 100 GHz, 200 GHz. The resonance peaks locate at the corresponding birefringence $\gamma_p/\pi$. Here $J_0/J_T=2$ and $P_{J0}=0$. 
}
\label{fig:6}
\vspace{-0.3cm}
\end{figure}

Due to the complexity of the analytical expressions for the {\em PM} response, we analyze numerically the effects of injection and birefringence on {\em PM}. 
As shown in Fig.~\ref{fig:6}(a), the {\em PM} resonant frequency and bandwidth increase only slightly ($<5$\%) with a three times larger injection. 
In contrast, the increase in birefringence significantly enhances the resonant frequency and bandwidth. 
Remarkably, the birefringence itself approximately determines the {\em PM} resonant frequency,  and the striking increase in the resonant frequency seen from Fig.~\ref{fig:6}(b)
is well described by 
$\omega_R^{PM} \approx \gamma_p$. Since birefringence larger than 200 GHz has been realized experimentally~\cite{Pusch2015:EL, Lindemann2019:N}, it can be employed to
overcome the bandwidth bottleneck~\cite{Hecht2016:N} of conventional {\em IM} ($\lesssim$35 GHz)~\cite{Haghighi2018:ISLC}. 
From the results in Fig.~\ref{fig:6}(b), guided by the room-temperature experiments on the highly-birferingent spin lasers~\cite{Lindemann2019:N}, we can see that the birefringence of 200 GHz
corresponds to the bandwidth of 300 GHz, about an order of magnitude larger than in the best conventional lasers~\cite{Haghighi2018:ISLC}, offering a promising approach for 
high-performance optical interconnect based on spin lasers. 

\begin{figure}[ht]
\centering
\includegraphics*[width=8.6cm]{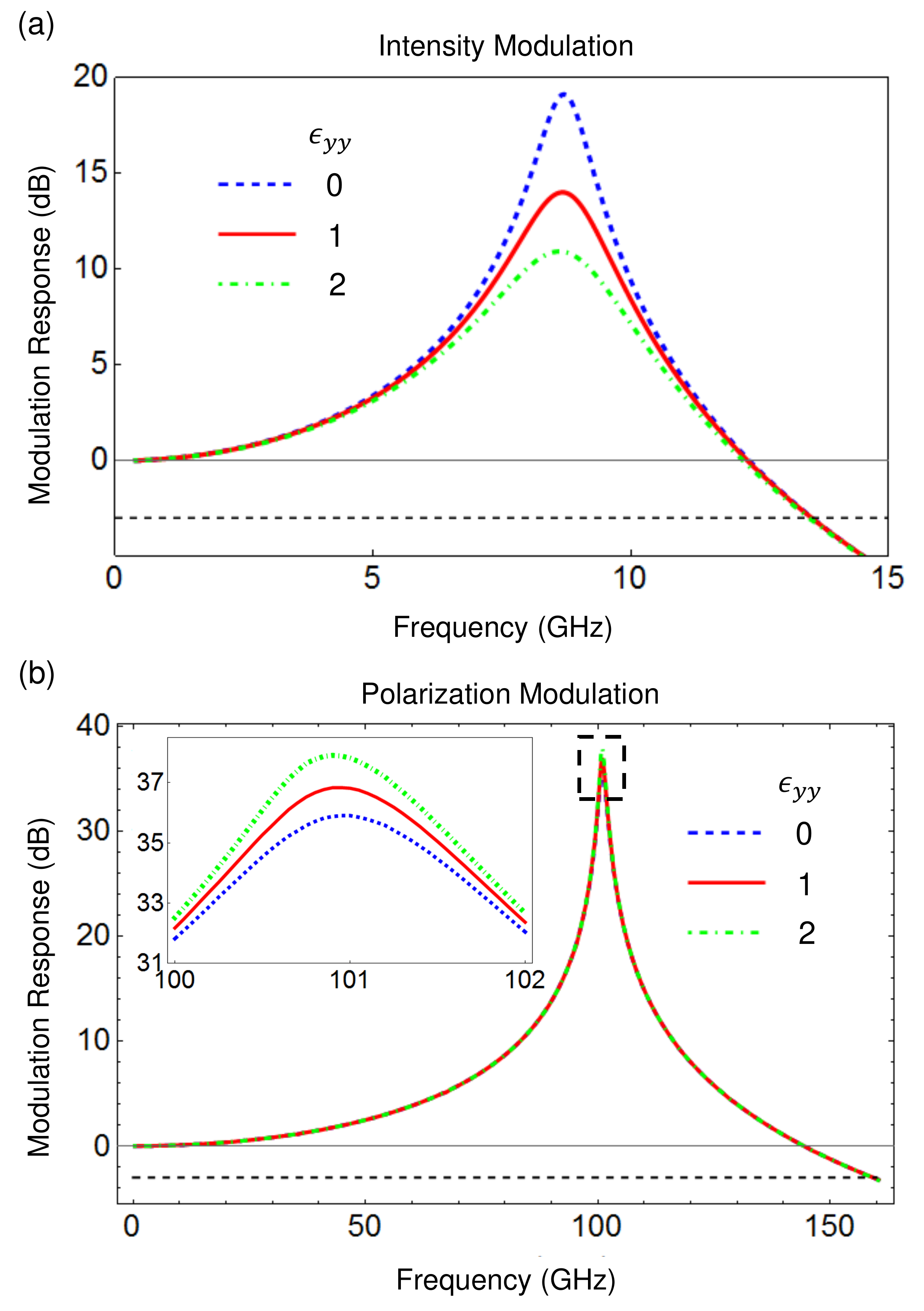}
\vspace{-0.5cm}
\caption{Self-saturation effects on intensity (a) and polarization modulation (b). (a) For intensity modulation, when self-saturation of the $y$ mode increases ($\epsilon_{yy} =0, 1, 2$), the response peak decreases, while the bandwidth is not affected. (b) For polarization modulation, the self-saturation effect is 
minor, only showing a slight increase in the response peak with larger self-saturation (inset). 
}
\label{fig:7}
\vspace{-0.3cm}
\end{figure}

A common strategy to increase the resonant frequency and  bandwidth in conventional lasers can be inferred from Eq.~(\ref{eq:3dB}) suggesting a desirable role of a
large-injection regime. However, depending on gain saturation, inevitable in semiconductor lasers~\cite{Lee1993:OQE}, which limits the intensity of 
emission with increasing injection, there is a detrimental impact on the modulation response and the increased power consumption. 

We illustrate in Fig.~\ref{fig:7} the effects of self-saturation, absent in SFM, on {\em IM} and {\em PM} response. For simplicity, we consider a case of $y$-mode lasing, i.e., $S_x \ll S_y$, which allows a focus on the saturation of the dominant $y$ mode, while the effect of $x$-mode saturation can be inferred analogously. For {\em IM}, the peak value of response is reduced with larger saturation 
$\epsilon_{yy}$, while the resonant frequency and bandwidth remain unchanged. The self-saturation effect on {\em PM} is much smaller and hardly noticeable, which can only been seen from 
the inset of Fig.~\ref{fig:7}(b). We see that the {\em PM} response is insensitive both to injection [Fig.~\ref{fig:6}(a)] and saturation, as it relies on the dynamics of polarization instead of intensity. 
Such distinct properties further make it a promising candidate for applications in low-energy ultrafast optical communication. Specifically, ultrafast operation in highly-birefringent spin lasers 
can be realized at low injection intensities, $J_T \lesssim J$, which has been recently demonstrated with electrically-tunable birefringence, even at elevated temperatures  
$\sim 70^\circ$ C~\cite{Lindemann2020:AIPA}. This could greatly reduce the power consumption, which is estimated to be an order of magnitude lower than in the state-of-the art conventional lasers~\cite{Lindemann2019:N,Moser2012:EL}. 

\section{V. Conclusions and Outlook}

The transparency of the developed intensity equations provides an intuitive description of intensity and polarization dynamics for both conventional and spin lasers.  
This approach, motivated by a popular spin-flip model~\cite{SanMiguel1995:PRA}, offers not only simpler calculations and  analytical results, but also a direct 
connection to widely-used rate equations~\cite{Chuang:2009,Coldren:2012} now including the missing description of optical anisotropies. 
While compared to the spin-flip model these intensity equations are obtained by eliminating the population difference between the spin-up and spin-down electrons,
 this approximation is accurately satisfied for spin lasers suitable for ultrafast operation and implementing optical interconnects~\cite{Lindemann2019:N,Yokota2018:APL,Lindemann2020:AIPA}. 

The introduced intensity equations overcome several limitations of the initial spin-flip model~\cite{SanMiguel1995:PRA}, which neglected gain saturation, particularly 
important for a large-injection regime, and assumed identical spin relaxation times of holes and electrons, despite characteristic times being typically several orders of magnitude 
shorter in holes~\cite{Zutic2004:RMP}. 
Instead, as relevant to most of the fabricated spin lasers, we have considered a vanishing spin relaxation time for holes. As shown within the 
generalized rate-equation description of spin lasers~\cite{Lee2014:APL}, this assumption can be relaxed to better describe GaN quantum well  spin 
lasers~\cite{FariaJunior2017:PRB}, where both electron and hole spin relaxation times are comparable~\cite{Brimont2009:JAP}, but have not been simultaneously 
considered in describing experiments~\cite{Bhattacharya2017:PRL}. 

Our findings on the modulation response reveal that for the intensity modulation, commonly used in conventional lasers, the corresponding resonant frequency and the bandwidth are independent of the experimentally demonstrated range of a linear birefringence. In contrast, for polarization modulation the resonant frequency, 
which can also give an estimate for the corresponding maximum bandwidth, grows linearly with the increase in such birefringence, to reach values largely exceeding 
the resonant frequency in fastest conventional lasers. There is a growing support that such improvements can be realized with different gain regions and cover a
wide range of the emitted light, from 850 nm to 1.55~$\mu$m~\cite{Lindemann2019:N,Lindemann2016:APL,Yokota2018:APL,Yokota2018:CLEO,Lindemann2020:AIPA,Lindemann2020:P}.

Presently, it is unclear what are the frequency limitations in the operation of spin lasers, for which both strain-enhanced birefringence and short spin relaxation times could 
help~\cite{FariaJunior2017:PRB,Lindemann2019:N}.
There are suggestions how the resonant frequency and the bandwidth could be further enhanced by choosing two-dimensional materials for the gain region and perhaps by employing magnetic proximity effects~\cite{Lindemann2019:N,Zutic2019:MT}. Instead of using pulsed ps optical spin injection (recall the approach from Fig.~3), it  would be desirable to seek alternative methods 
for modulation of the carrier spin polarization and consider phenomena that were previously not studied in the context of spin lasers. For example, using ultrafast 
demagnetization~\cite{Battiato2010:PRL,Kampfrath2013:NN}, ultrasfast magnetization reversal~\cite{Garzon2008:PRB, Kirilyuk:2019}, or ultrafast modulation of spin and optical polarization using bound states in quantum  wells~\cite{Rozhansky2020:PRB,Mantsevich2019:PRB}. Gate-controlled reversal of helicity was predicted in two-dimensional topological materials~\cite{Xu2020:PRL}, while electrical injection from iron GaAs-based light-emitting diode was demonstrated to support helicity switching at room 
temperature~\cite{Nishizawa2017:PNAS,Nishizawa2018:APE,Munekata2020:SPIE}.

While our focus was on vertical cavity surface emitting lasers (VCSELs)~\cite{Michalzik:2013}, typically used to implement spin laser, it would be interesting 
to consider if  these intensity equations can also complement the studies of vertical external cavity surface emitting lasers (VECSELs)~\cite{Frougier2013:APL,Frougier2015:OE}.  
They have complementary advantages to VCSELs and
having an external cavity may offer an additional control optical anisotropy, including birefringence, as well as incorporate magnetic elements close to the gain region
for efficient electrical spin injection~\cite{Frougier2015:OE,Alouini2018:OE}. Efforts to obtain an efficient room-temperature electrical injection in semiconductors 
with  perpendicular magnetic anisotropy of the spin injector~\cite{Sinsarp2007:JJAP,Hovel2008:APLa,Zarpellon2012:PRB}
could be extended in spin lasers to remove the need to use an external magnetic field to align the magnetization out-of-plane, 
consistent with the usual optical selection rules~\cite{Zutic2004:RMP}.

In addition to the relevance of spin lasers as emerging room-temperature spintronic devices with operation principles not limited by magnetoresistive 
effects~\cite{Zutic2004:RMP,Tsymbal:2019,Zutic2020:SSC,Hirohata2020:JMMM,Lin2019:NE}, the studied intensity equations could also be helpful
in exploring other device concepts. For example, an earlier work on rate equations~\cite{Lee2010:APL,Lee2012:PRB} was helpful to motivate electrical 
spin interconnects~\cite{Dery2011:APL,Zutic2011:NM,Zutic2019:MT} and phonon lasers~\cite{Khaetskii2013:PRL}, an acoustic analog 
of lasers which also shares properties with spin-controlled nanomechanical resonators~\cite{Stadler2014:PRL,Mantovani2019:PRB}.  

\section*{Acknowledgments}
This work has been supported by the NSF ECCS-1810266. We thank N. C. Gerhardt, M. Lindemann, M. R. Hofmann,  R. Michalzik, and T.  Pusch for valuable discussions.

\appendix
\section{APPENDIX}
\renewcommand{\theequation}{A\arabic{equation}}
\setcounter{equation}{0}

The quantities in the spin-flip model (SFM) equations~\cite{SanMiguel1995:PRA} are usually studied in the dimensionless form making
it important to describe how they are normalized and simplify their relation to other rate-equation description of lasers. Specifically, the
quantities in SFM 
have been normalized as 
\begin{figure}[t]
\centering
\includegraphics*[width=8.6cm]{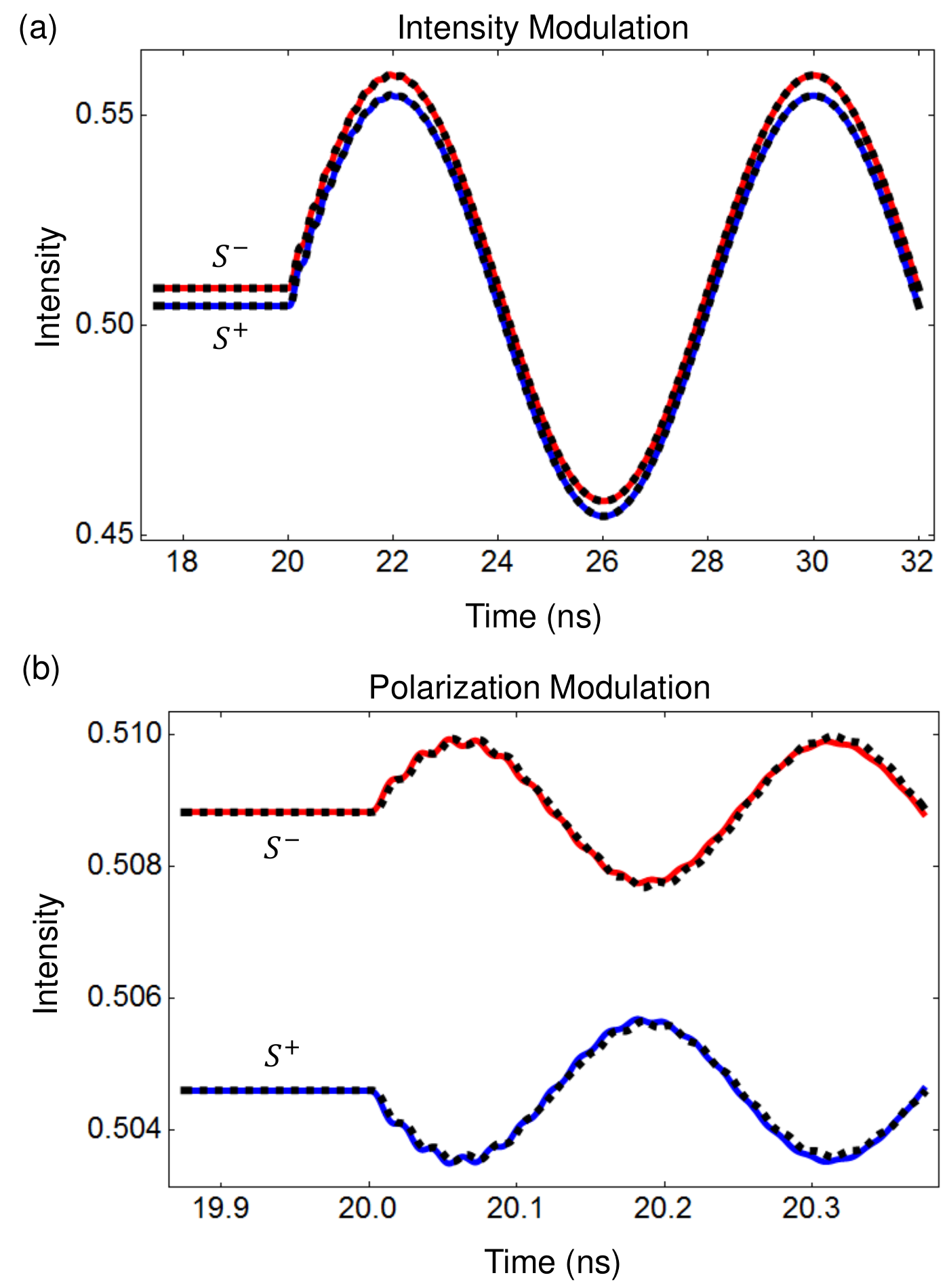}
\vspace{-0.5cm}
\caption{Comparison of time evolution of polarization-resolved intensities $S^{\pm}$ between intensity equations (solid) and SFM (dashed) under intensity modulation (a) and polarization modulation (b). Before modulations turned on at time $t=20$ ns, the injection is constant with a spin polarization $P_{J0}=0.1$.  Parameters: $J_0=2$, $\gamma_p/\pi = 50$ GHz, $\gamma_s = 450$ GHz. 
}
\label{Appendix_fig:1}
\vspace{-0.3cm}
\end{figure}
\begin{eqnarray}
E_{\pm} &=&\frac{F_{\pm}}{\sqrt{S_{2J_{T}}}} , \\
N &=& \frac{N_+ + N_- - N_\mathrm{tran}}{N_{T} - N_\mathrm{tran}}, \\
n &=& \frac{N_- - N_+}{N_{T} - N_\mathrm{tran}},
\end{eqnarray}
where $F_{\pm}$ are the slowly varying amplitudes of the  helicity-resolved components of the electric field, $S_{2J_{T}}$ is the steady-state light intensity at twice the threshold injection $2J_{T}$,  $N_{\pm}$ are the numbers of spin-up and spin-down electrons, $N_{T}$ and $N_\mathrm{tran}$ are the numbers of electrons at the threshold and transparency, respectively. 
The injection $J$ has been normalized with respect to threshold injection $J_T$. We have assumed $\gamma_a\ll1/\tau_{ph}$ in the above normalizations. 

\begin{figure}[ht]
\centering
\includegraphics*[width=8.6cm]{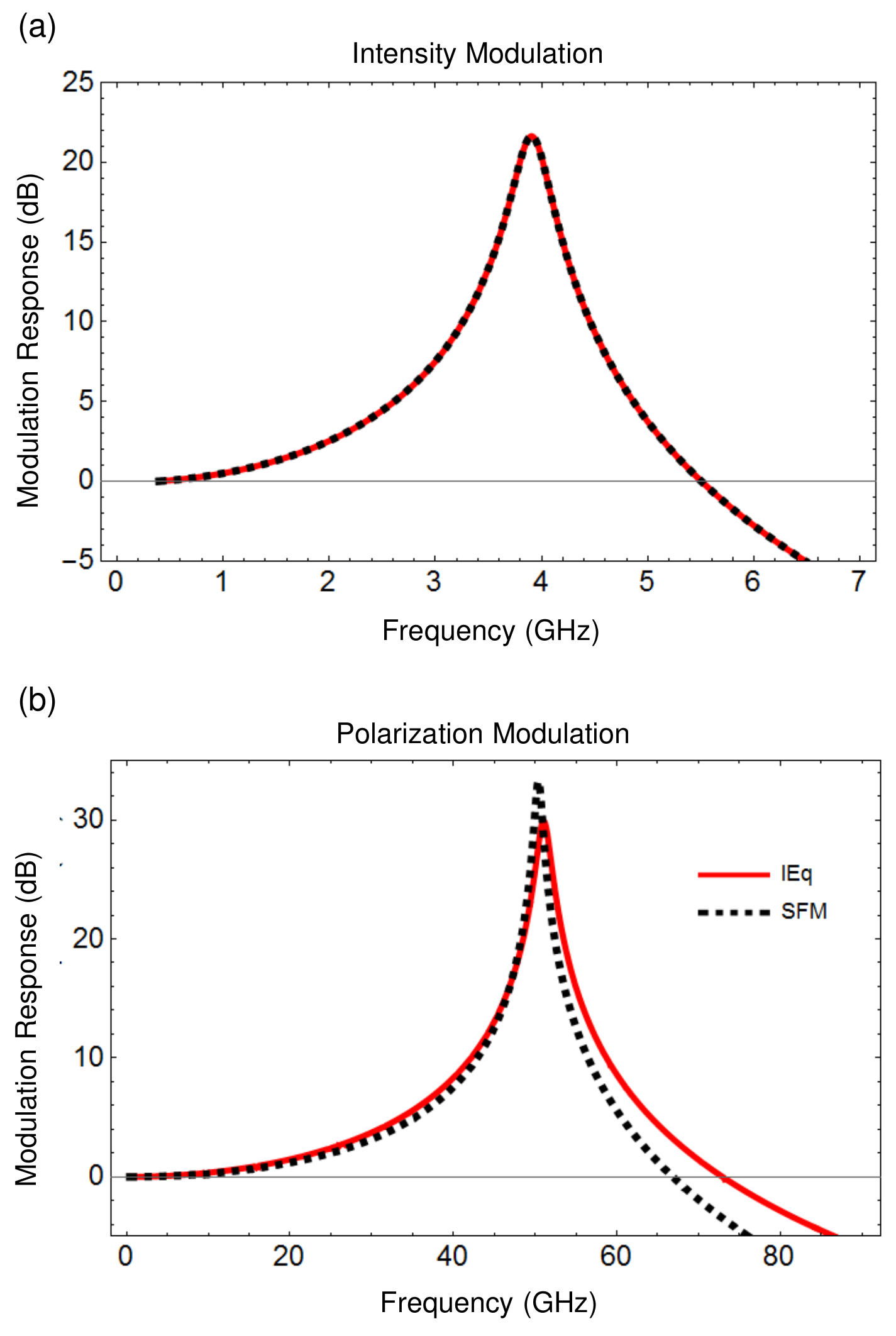}
\vspace{-0.5cm}
\caption{Comparison of the response under intensity modulation (a) and polarization modulation (b) between intensity equations (solid) and SFM (dashed). Parameters: $J_0=2$,  $P_{J0}=0$, $\gamma_p/\pi = 50$ GHz, $\gamma_s = 450$ GHz. 
}
\label{Appendix_fig:2}
\vspace{-0.3cm}
\end{figure}

To verify the validity of intensity equations, we compare the numerical results from the intensity equations and SFM. In Fig.~\ref{Appendix_fig:1}, we show a comparison of the time evolution of helicity-resolved intensities $S^{\pm}$ between intensity equations and SFM under {\em IM} and {\em PM}, respectively. We see that the agreement in the time evolution is excellent, with only a minor deviation for {\em PM}. The comparison of modulation response is illustrated in Fig.~\ref{Appendix_fig:2}, which shows a good overall agreement, with only a small discrepancy in the 
{\em PM} response. 

\bibliographystyle{apsrev4-1}

\begin{thebibliography}{99}
\bibitem{const}
This is not essential but simplifies our analytical results.
\end{thebibliography}

\end{document}